# Magnetic field induced effects in the high source-drain bias current of weakly coupled vertical quantum dot molecules

D G Austing,[1,2][*] C Payette,[1,2] G Yu,[1] J A Gupta,[1]

[1]*Institute for Microstructural Sciences M50, National Research Council of Canada, 1200 Montreal Road, Ottawa, Ontario K1A 0R6, Canada*

[2]*McGill University, Department of Physics, Ernest Rutherford Physics Building, 3600 rue University, Montréal, Quebec H3A 2T8, Canada*

**Abstract**

We report on the basic properties of recently observed magnetic field resonance, induced time dependent oscillation, and hysteresis effects in the current flowing through two weakly coupled vertical quantum dots at high source-drain bias (up to a few tens of mV). These effects bare some similarity to those reported in the N=2 spin-blockade regime, usually for weak in-plane magnetic field, of quantum dot molecules and attributed to hyperfine coupling, *except* here the measurements are conducted outside of the spin-blockade regime and the out-of-plane magnetic field is up to ~6 T.

## 1. Introduction

Several intriguing effects, including current switching and hysteresis, and slow current oscillations or fluctuations on a long time scale (seconds and minutes), all attributed to electron spin–nuclear spin (hyperfine) coupling, have been reported in the low bias, low magnetic field (mostly below 1 T), two- or effective two-electron spin-blockade regime [1] for vertical double dots, lateral double dots, and InAs nanowire double dots [2-9]. Although at first sight hyperfine coupling appears potentially ruinous for quantum information applications involving III-V quantum dots, hyperfine effects have in fact turned out to be advantageous for recent key one- and two-qubit operations [5,7], and nuclear quantum memory has also been proposed [10]. The above mentioned experiments have led to many interesting theoretical works [11-23], although several features remain to be fully explained.

We outline some basic properties of magnetic (B-) field induced resonance, time dependent oscillation, and hysteresis effects we have seen in the current flowing through two vertically coupled vertical quantum dots at *high source-drain bias* (up to a few tens of mV and thus outside of the N=2 spin blockade regime), and *high magnetic field* (up ~6 T).

## 2. Device Structure and Measurement Regime

The device is a sub-micron gated AlGaAs/InGaAs/AlGaAs/InGaAs/AlGaAs triple barrier structure for which the tunnel coupling between the two quantum dots is very weak (<0.1 meV) [1]. Similar looking effects, due to hyperfine coupling, have been reported in the low bias two-electron spin-blockade regime of vertical double dots for B<1 T [2], and of lateral double dots for B<0.2 T [4], when the B-field is applied *parallel* to the growth plane of the structure. The regime we study here is for high bias and higher B-field (usually in the range

[*] Corresponding author. Tel.: +1-613-991-9989; fax: +1-613-990-0202; e-mail: guy.austing@nrc-cnrc.gc.ca.

of 1 to 6 T) applied *perpendicular* to the growth plane of the structure.

## 3. Discussion

The 0 T differential conductance in the source drain bias voltage-gate voltage ($V_{sd}$-$V_g$) plane is given in Fig. 1. The resonance and hysteresis effects we observe in the DC current occur throughout a wide region, roughly but not exclusively inside the boxed region, on sweeping the B-field up and down. This region extends to high bias, and is away from the (shaded) chevron shaped region of suppressed current due to spin-blockade that lies to one side of the N=2 Coulomb blockade region [1].

Figure 2 shows typical up-sweep and down-sweep traces at three different (high) source drain biases for a fixed gate voltage. The sweep rate is 125 mT/min so each up-down sweep takes ~1 hour. The up-sweep and down-sweep traces *separately* exhibit sharp current resonances (amplitude up to ~5 pA) and "steps". There is also prominent hysteresis between up-sweep and down-sweep traces. The resonances and hysteresis are reproducible at fixed source drain bias, gate voltage, B-field sweep rate and temperature (visible up to at least 4 K), and are pronounced between 1 and 6 T [24].

Up-sweep and down-sweep traces near one particular sharp (~25 mT wide) resonance at ~4 T are given in Fig. 3 (a). The sweep rate is a very slow ~2 mT/min and the extent of the hysteresis between the up-trace and down-trace is 5-10 mT. A 13 minute time segment trace in Fig. 3 (b) reveals quasi-periodic multi-period current oscillations at one particular fixed B-field position within the resonance with a strong long period component (~150 sec) and weaker shorter period components (few tens of seconds and shorter). The maximum amplitude of the oscillation is about 2.5 pA, whereas to the left and right of the resonance the current is "flat" with a noise-level of a few tens of fA.

The effects described have been seen in four different double dot devices. Although the exact details vary from device-to-device, the extent of the hysteresis in the position of corresponding features in the up-sweep and down-sweep traces is typically between a few tens of mT and a couple hundred mT for sweep rates respectively of 15 and 300 mT/min. Furthermore, "slow" current oscillations and fluctuations are generally observed on the timescale of seconds to many tens of seconds on entering the sharp (typically a few tens of mT or less wide) current resonances [24].

These observations suggest a hyperfine related origin [2,4] even though the effects occur well outside the low bias N=2 spin-blockade region. A model for hyperfine coupling in the complex non-linear high bias regime, also requiring knowledge of B-field induced (orbital and Zeeman) crossings of N>2 double dot states, including dot-dot detuning, is currently lacking. Some clues may be present in recent experiments involving spin-blockade for (even) N>2 [8-9,25].

## Acknowledgements

We are grateful for the assistance of A. Bezinger, D. Roth, and M. Malloy for micro-fabrication. CP is funded by DGA's NSERC Discovery Grant. Part of this work is supported by the DARPA-QUIST program (DAAD19-01-1-0659).

[18] D. Klauser, W. A. Coish, and D. Loss, Phys. Rev. B 73 (2006) 205302.
[19] G. Ramon and X. Hu, Phys. Rev. B 75 (2007) 161301.
[20] M. Stopa, and C. M. Marcus, arXiv:cond-mat/0604008.
[21] D. Klauser, W. A. Coish, and D. Loss, arXiv:cond-mat/0604252.
[22] J. Inarrea, G. Platero, and A.H. MacDonald, arXiv:cond-mat/0609323.
[23] M. S. Rudner, and L. S. Levitov, arXiv:0705.2177.
[24] D. G. Austing, et al., (2007) unpublished.
[25] A. C. Johnson et al., Phys. Rev. B 72 (2005) 165308.

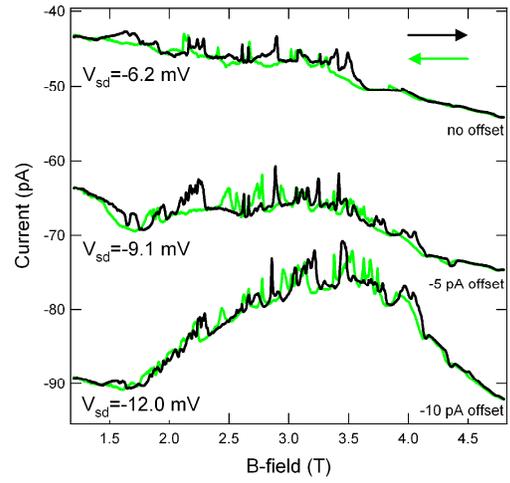

Fig. 2 Up-sweep (black) and down-sweep (green) traces at three different biases, and at gate voltage -1.04 V. The sweep rate is 125 mT/min. For clarity, the middle and lower trace pairs have been offset.

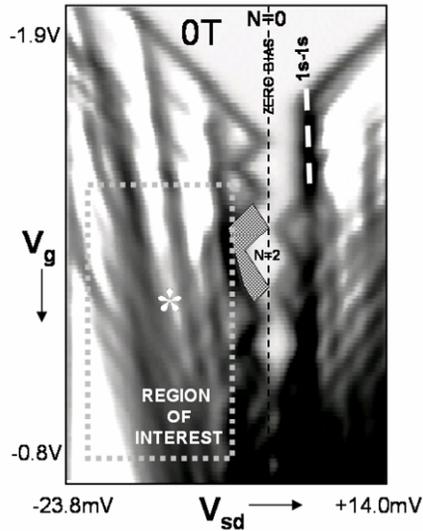

Fig. 1 0 T differential conductance in the $V_{sd}$-$V_g$ plane. Black, grey, and white respectively corresponds to positive, zero, and negative conductance. The 1s-1s resonance occurs in the opposite bias direction to that of the boxed region of interest and not at zero bias because of energy mismatch between the two dots.

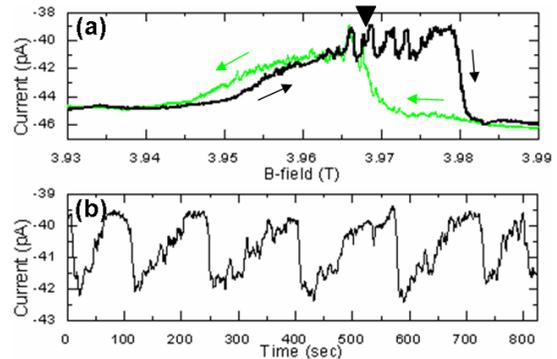

Fig. 3 (a) Up-sweep (black) and down-sweep (green) traces at a bias of -12 mV and gate voltage -1.22 V. This fixed ($V_{sd}$,$V_g$) point is marked with as asterisk in Fig. 1. The sweep rate is ~2 mT/min. (b) A 13 minute time segment trace of the current at the fixed B-field marked by the triangle in (a).